\documentclass[3p]{elsarticle}

\usepackage{amsmath,amssymb,graphicx, amsthm}

\usepackage{tikz}
\usepackage{mathdots}
\usepackage{yhmath}
\usepackage{cancel}
\usepackage{siunitx}
\usepackage{array}
\usepackage{multirow}
\usepackage{gensymb}
\usepackage{tabularx}
\usepackage{booktabs}
\usetikzlibrary{fadings}
\usepackage{standalone}
\usepackage[ruled,vlined]{algorithm2e}

\usepackage{longtable}
\usepackage{verbatim}
\usepackage{footnote}
\usepackage{hyperref}
 
\usepackage{color}

\usepackage{natbib}
\setcitestyle{authoryear,open={(},close={)}}

\usepackage{amsmath}
\DeclareMathOperator*{\argmax}{arg\,max}

\title{
Interpretable  brain age prediction using 
linear latent variable models of 
functional connectivity
}

\author[1,8]{Ricardo Pio Monti} 
\ead{r.monti@ucl.ac.uk}
\author[2]{Alex Gibberd}
\ead{a.gibberd@lancaster.ac.uk}
\author[3]{Sandipan Roy}
\ead{s.roy@bath.ac.uk}
\author[3]{Matt Nunes}
\ead{m.nunes@bath.ac.uk}
\author[4,5]{Romy Lorenz}
\author[6]{Robert Leech}
\ead{r.leech@kcl.ac.uk}
\author[7]{Takeshi Ogawa}
\author[8]{Motoaki Kawanabe}
\author[10,9,1]{Aapo Hyv\"{a}rinen}
\ead{}

\address[1]{Gatsby Computational Neuroscience Unit, University College London, London, UK}
\address[2]{Department of Mathematics \& Statistics, Lancaster University, Bailrigg, UK}
\address[3]{Department of Mathematics, University of Bath, Bath, UK}
\address[4]{MRC Cognition and Brain Sciences Unit, University of Cambridge, Cambridge, UK}
\address[5]{Max-Planck Institute for Human Cognitive and Brain Sciences, Leipzig, Germany}
\address[6]{Centre for Neuroimaging Science, Kings College London, London, UK}
\address[7]{Advanced Telecommunications Research Institute International, Cognitive Mechanisms Laboratories, Kyoto, Japan}
\address[8]{Brain Information Communication Research Laboratory Group, Advanced Telecommunications Research Institute International (ATR), Kyoto, Japan}
\address[9]{Department of Computer Science and HIIT, University of Helsinki, Helsinki, Finland}
\address[10]{INRIA, Paris-Saclay University, Paris, France}

\begin{document}

\begin{abstract}

Neuroimaging-driven prediction of brain age, defined as the predicted biological age of a subject 
using only brain imaging data, is an exciting avenue of research.
In this work we
seek to build models of brain age based on functional connectivity while 
prioritizing model 
interpretability and understanding. This way, the models serve to both provide accurate estimates of brain age 
as well as 
allow us to investigate changes in functional connectivity which occur during the ageing process. 
The methods proposed in this work consist of a two-step procedure: first, 
linear latent variable models, such as PCA and its extensions,  are employed to learn reproducible 
functional connectivity
networks present across 
a cohort of subjects. The activity within each 
network is subsequently employed as a  feature in a linear regression model to predict
brain age. 
The proposed framework is employed on the data from the CamCAN repository
and the inferred brain age models are further demonstrated to 
generalize using data from two open-access repositories: the Human Connectome Project and the ATR Wide-Age-Range.

\end{abstract}

\maketitle

\section{Introduction}




The human brain changes during the lifespan of an adult, resulting in 
robust and reproducible changes in structure and function \citep{raz2006differential, lim2013preferential}.
Moreover, there is reason to hypothesize that deviations from the typical brain ageing trajectory 
may reflect latent neuropathological influences \citep{cole2018brain}, 
serving to motivate 
further research into developing 
reliable biomarkers derived from brain imaging data. 
Such biomarkers could be fundamental in order to better understand and combat 
age-associated neurodegenerative diseases.  To date, early studies 
have shown success in the context of traumatic brain injury \citep{cole2015prediction} and  schizophrenia 
\citep{koutsouleris2013accelerated}.

Due to the significant potential benefits associated with brain-imaging driven biomarkers 
for age, there have been 
many statistical models proposed for healthy brain ageing. 
These models vary in complexity as well as in the class of 
neuroimaging data employed.  
One of the earliest demonstrations was that of
\cite{good2001voxel}, who employed voxel-based morphometry to demonstrate the structural changes which occur during healthy ageing. 
More recently,
a wide range of sophisticated machine learning methods have been employed 
\citep{franke2013advanced, lancaster2018bayesian, smith2019estimation}. 
\cite{cole2015prediction} employed Gaussian process regression to predict the biological age of 
subjects using structural neuroimaging data, demonstrating that such a model was able to accurately predict
brain age. The resulting model was subsequently applied to subjects with traumatic brain injury (TBI),
where the associated residuals (difference between predicted and true biological age)
were shown to be significantly larger for subjects with TBI as compared with healthy subjects; 
the associated model 
consistently predicted subjects with TBI to be \textit{older}, possibly a result of accelerated atrophy. 
This work was further extended by \cite{cole2017predicting_CNN}, 
who employed convolutional neural networks
to obtain improved performance. In related work, \cite{franke2010estimating} employ kernel regression
with an application to the early identification of Alzheimer's disease. 

While the vast majority of the literature has employed
structural imaging modalities, there are also numerous examples of
where 
functional imaging has been utilized. 
A pertinent example is
\cite{dosenbach2010prediction}, 
who employ resting-state fMRI together with support vector machines (SVMs) in order to 
accurately classify subjects as being either children 
(ages 7-11 years old) or adults (ages 24-30 years old). 
Furthermore, they observe an overall decrease in 
network connectivity as subjects mature. 
In  related work, \cite{geerlings2012reduced} identify ageing-driven changes in functional connectivity, highlighting 
decreased connectivity within the default mode network and the somatomotor network. 
Subsequently, 
\cite{geerligs2014brain} categorized the changes in functional connectivity that 
occur with healthy ageing in terms of various network measures.

More generally, the study of functional connectivity is itself an exciting avenue of modern neuroscientific research which
has shown great potential for improving our understanding of the human brain function and architecture \citep{sporns2012discovering}. By way of example, changes in functional connectivity have been related to various  neuropathologies such as  Parkinson's disease \citep{Wu2009} and Alzheimer’s \citep{Damoiseaux2011}
as well as conditions such as Autism \citep{Cherkassky2006}. 
Recently, the changes in functional connectivity induced by ageing have begun to be studied.
Initial studies have reported 
significant differences in the connectivity between younger and older subjects using resting-state fMRI \citep{geerligs2014brain}.
Moreover, results appear to suggest there are important changes that occur in the connectivity not just between regions but also
at the level of entire networks.
%
However, despite recent advances, a holistic understanding of the relationship between healthy ageing and 
the associated changes in functional connectivity is still missing. 

In this work we seek to build robust models of brain age based on the  functional connectivity of 
individuals. This serves to combine the two prominent avenues of neuroscientific research:
brain age prediction and analysis of functional connectivity.  
In particular,  the methods presented in this work have two principal objectives:
\begin{enumerate}
    \item To demonstrate that measures of functional connectivity can reliably be employed as features in
    machine learning models of brain age. To this end we build and validate models using three large open-source datasets: 
    the Cambridge Center for Ageing and Neuroscience (CamCAN), 
    the Human Connectome Project (HCP) and 
    the ATR Wide-Age-Range datasets.
    \item We further wish to interpret and inspect the proposed models in order to gain further insights into the 
    changes in functional connectivity associated with ageing. This calls for the use of parsimonious and simple predictive models together 
    with features whose relationship with functional conncetivity is clearly understood.
\end{enumerate}

Throughout this paper, we put forward the thesis that for the potential impact of functional connectivity assessment to be met (i.e., in terms of developing powerful biomarkers) the research community needs to develop robust methods for data-analysis which can combine both supervised and unsupervised models of functional connectivity analysis. 
Instead of tweaking existing statistical methods, it is imperative to
develop methods which 
are intuitive, interpretable, and insightful from a neurophysiological perspective.
Such models must utilise as much experimental information as possible
in order to investigate the 
factors which affect functional connectivity.

To further  motivate our thesis, one should consider that most experiments to date operate on data from a single laboratory, or class of experiment which limits the generality of any obtained results. 
Such concerns have been recently recognised, particularly within the 
context of brain ageing
\citep{geerligs2015state, geerligs2017challenges},
and have given 
rise to multi-laboratory collaborations with data-sharing becoming more common.
However, it is still highly unlikely that all subject features (and how these are measured) will be comparable across different experimental environments. Thus while data-sharing has seen much progress, it could be argued that the impact of these endeavours is still to come, 
and to achieve this, we need to develop methods which can combine information from across disparate, but informative experiments.

\begin{figure}[t!]
    \centering
    \includegraphics[width=\textwidth]{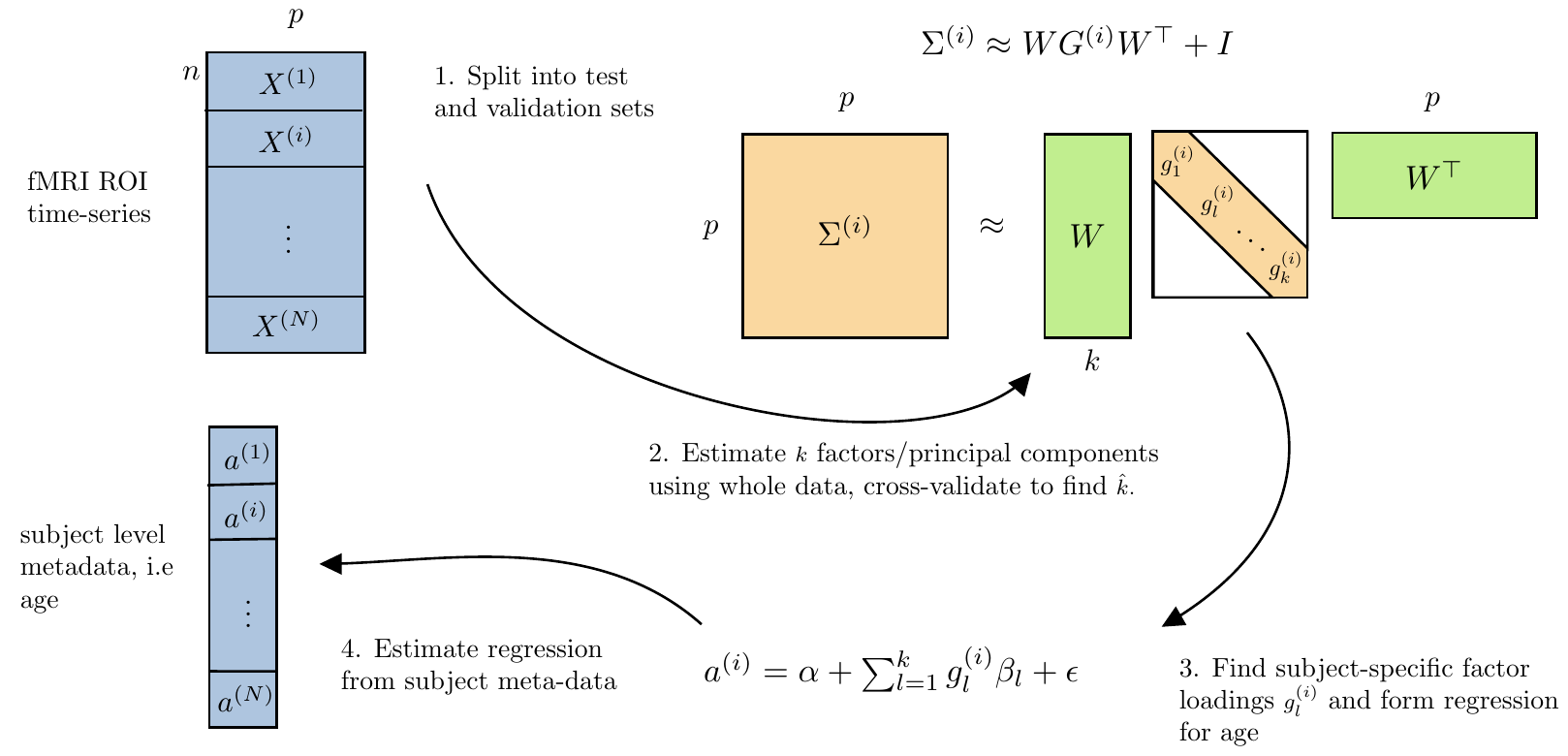}
    \caption{Pipeline for estimating networks, factor loadings, and predictive model for biological brain age.
   	Inferred factors $W\in \mathbb{R}^{p\times k}$ describe networks which are reproducible
    across the whole population, the subject-specific factor loadings $g_l^{(i)}$ are then used to predict brain age. Once the factor loadings are estimated as above, using one experimental data-set (we use CamCAN data in our experiments), we can then assess how these factors perform for brain age prediction on completely held-out data-sets; we demonstrate how the model generalizes well using HCP and ATR Wide-Age-Range datasets. }
    \label{fig:pipeline}
\end{figure}

To this end we proceed in a two-step framework. First, we seek to learn robust features which summarize
properties of functional connectivity across a cohort of subjects in an unsupervised manner.
Due to our focus on interpretability, we focus on linear latent variable models, such as principal component analysis (PCA) and
its generalizations. 
The benefit of employing latent variable models such as PCA is that we may interpret the 
latent variables in terms of activity within functional connectivity 
networks, as  proposed by \cite{Leonardi2013} (see also Figure \ref{fig:PCAextensions} below). 
%
Second, once features have been obtained in an unsupervised manner, they are subsequently used to predict brain age 
using standard linear regression models. We deliberately restrict ourselves to simple linear classifiers as they can be 
easily interrogated, allowing us to explicitly understand how each feature contributes to the predicted brain age. 
An overview of our two-stage approach is provided in Figure \ref{fig:pipeline}.

The remainder of this manuscript is organized as follows: 
in Section \ref{sec::methods} we first review linear latent variable models and their implications for 
functional connectivity analysis. We then present the proposed two-step procedure. 
Experimental results, studying synthetic as well as real resting-state fMRI data, are presented in Section \ref{sec:ExpRes}. 

\section{Methods}
\label{sec::methods}
We focus our analysis on resting-state fMRI time series 
data which is collected across a cohort of $N$ subjects. 
For the $i$th subject, it is assumed we have access to 
fMRI measurements over $p$ fixed regions of interest, denoted by  $X^{(i)} \in \mathbb{R}^{ p}$, 
as well as the subjects age, $a^{(i)} \in \mathbb{R}_+$. 
Throughout this work
we approximately model the fMRI data for each subject 
with a stationary multivariate Gaussian distribution,  
$ X^{(i)} \sim \mathcal{N} (0, \Sigma^{(i)})$, where $\Sigma^{(i)}$ denotes the covariance for subject $i$.
Each entry in $\Sigma^{(i)}$ denotes the  covariance between any pair of regions, which serves to 
define a measure of the functional connectivity \citep{smith2012future}. As such, it follows that $\Sigma^{(i)}$ encodes 
a functional connectivity network over $p$ regions where edges encode the marginal dependence structure. 

The goal of the proposed methods is to learn interpretable and robust models to predict the 
biological age, $a^{(i)}$, of subjects given information relating only to their functional connectivity. 
To achieve this, we 
propose a two-step framework.
Our approach first employs linear latent variable models in order to model high-dimensional connectivity 
matrices using a reduced number of latent variables. We interpret such variables 
as corresponding to functional connectivity networks,
allowing us to describe patterns in connectivity as being composed of 
various distinct networks. 
We note that such a two-step approach has previously been employed in the context of brain age prediction 
\citep{franke2010estimating, smith2019estimation}. 
However, as far as we are aware, this is the first work to directly interpret the role of linear latent variable 
models, such as PCA, as learning the relevant functional networks. This work thereby provides a clear motivation and 
interpretation for such a two-stage strategy.

In Section \ref{linearLVMreview} we discuss the various latent variable models employed, and highlight how introducing assumptions 
such as non-negativity can help further improve interpretability of results. We also discuss theoretical benefits associated with such 
assumptions. 
We then discuss the how the features (i.e., functional networks) inferred by the latent variable models may be used to build 
linear models for brain age. 

\subsection{Linear latent variable models for functional connectivity: PCA and its extensions}
\label{linearLVMreview}

In this section we outline the linear latent variable models employed in the unsupervised learning 
stage of the proposed framework. 
We begin by discussing principal component analysis (PCA), a well-established technique for dimensionality reduction
\citep{jolliffe2011principal}. The common derivation for PCA poses it as an optimization problem seeking to learn the 
linear projection which maximizes explained variance within the projected space \citep{hotelling1933analysis}. 
However, 
PCA can also be derived as inference under a simple linear latent variable model, which posits 
that observations $X^{(i)} \in \mathbb{R}^p$ are generated as a linear projection from low-dimensional latent variables, $Z^{(i)} \in \mathbb{R}^k$
\citep{harman1960modern}.
%
When both observations and latent variables are taken to follow a multivariate Gaussian distributions we obtain the following generative model for observed data:
\begin{align}
    \label{pca_gen_1}
    Z^{(i)} &\sim \mathcal{N}(0, G^{(i)}) \\
    \label{pca_gen_2}
    X^{(i)} | Z^{(i)} =z^{(i)} &\sim  \mathcal{N}( W z^{(i)}, v^{(i)} I)
\end{align}
where $G^{(i)}\in \mathbb{R}^{k \times k}$ is a diagonal matrix and $v^{(i)} \in \mathbb{R}_+$ denotes measurement noise. 
Equations (\ref{pca_gen_1}) and (\ref{pca_gen_2}) serve to highlight how PCA can be seen as a low-rank model for the covariance matrix;
by marginalizing over latent variables we obtain: 
\begin{equation}
    \label{linLVM_covModel}
    \Sigma^{(i)} = W G^{(i)} W^T + v^{(i)} I,
\end{equation}
implying that the loading matrix, $W$, captures low-rank covariance structure. 
Learning the associated loading matrix, $W$, proceeds via maximizing the  log-likelihood over observations across all
$N$ subjects:
\begin{equation}
    \mathcal{L} = \sum_{i=1}^N p \log 2 \pi + \log \mbox{det} ~\Sigma^{(i)} + \mbox{tr} \left ( {\Sigma^{(i)}}^{-1} K^{(i)} \right ),
\end{equation}
where $\Sigma^{(i)}$ is as defined in equation (\ref{linLVM_covModel}) and $K^{(i)}$ denotes the sample covariance matrix for the $i$th subject. 
In the context PCA, the maximization is performed subject to the constraint that $W$ be orthonormal:
\begin{equation}
    \label{PCA_W}
    \hat W = \argmax_{W: W^T W = I } \left \{  \mathcal{L} \right \},
\end{equation}
and a closed-form solution is obtained via eigendecomposition. 

Following \cite{Leonardi2013} it is possible to interpret 
each column of $W$ as encoding functional networks or ``eigenconnectivities". While the loading matrix, $W$, is shared across all subjects, 
each diagonal entry of $G^{(i)}$ denotes the extent to which
the associated network is expressed in subject $i$. This allows us to study  connectivity as being composed of 
various distinct networks, resulting in 
significant benefits from the perspective of interpretability. 
We can further unpack equation 
(\ref{linLVM_covModel}) as follows (see also Figure \ref{fig:PCAextensions} below):
\begin{equation}
        \label{linLVM_covModel_exapanded}
    \Sigma^{(i)} = \sum_{j=1}^k g^{(i)}_j W_j W_j^T  + v^{(i)} I,
\end{equation}
where $W_j$ denotes the $j$th column of $W$ and we write $g^{(i)}_j$ to denote the $j$th diagonal entry 
of the matrix $G^{(i)} \in \mathbb{R}^{k \times k}$. As such, we may interpret 
each $W_j$ as encoding the $j$th 
network and $g_j^{(i)}$ as a measure of activity within the corresponding 
network in the $i$th subject.

There exists several extensions to the model described in equations (\ref{pca_gen_1}) and (\ref{pca_gen_2}),
the prime example being factor analysis which allows the variances in equation (\ref{pca_gen_2}) to vary across dimensions. 
Recently, several extensions have been proposed 
where constraints such as non-negativity are introduced with the goal of improving the 
interpretability of results \citep{Zass2007nonnegative, sigg2008expectation, hirayama2016characterizing}.
The motivation behind such methods stems from the fact that interpreting and visualizing 
PCA-based 
networks becomes very challenging, particularly in high-dimensions. 
Challenges arise from the fact that each principal component will correspond to a weighted sum of BOLD activities across 
all observed regions. As such, it is often difficult to identify which regions are the principal contributors to a certain 
principal component (and hence functional network) without applying ad-hoc post analysis. Furthermore, it is possible that some entries in
the principal components may be negative,
which further complicates the interpretation from the perspective of functional connectivity analysis.

The aforementioned issues can be mitigated via the introduction of non-negativity constraints on the loading matrix, $W$. 
This ensures that each principal component corresponds only to a weighted \textit{positive} sum of activity over 
all brain regions. As such, the principal component can be directly interpreted as the contribution of each region to each
functional
network. Furthermore, the introduction of non-negativity will often yield sparsity in the sense that many of the entries of 
the principal components will be exactly zero \citep{sigg2008expectation}. It follows that such sparsity further facilitates the 
interpretation of the corresponding 
networks. 
From an optimization perspective, 
the loading matrix is inferred by maximizing the original log-likelihood objective, with the additional non-negativity constraint:
\begin{equation}
    \label{NNPCA_W}
    \hat W = \argmax_{W: W \geq 0 } \left \{  \mathcal{L} \right \}. 
\end{equation}

It is important to note that the orthonormality constraint has been dropped in
equation (\ref{NNPCA_W}), making the associated optimization problem less challenging. 
However, the combination of 
non-negativity and orthonormality, as  enforced in
\cite{Monti2018}, 
leads to several desirable properties. 
First, the loading matrix $W$ has at most one non-zero entry per row. This implies that we may 
interpret the columns of $W$ as encoding membership to $k$ non-overlapping networks or clusters.  
Another very important benefit 
of introducing non-negativity and orthonormality constraints
is that the matrix $W$ is uniquely defined and identifiable.
This is not the case in standard 
factor analytic models, where $W$ is only identifiable up to 
an arbitrary rotation \citep{harman1960modern, bishop2006pattern}.
Given that throughout this work we will directly interpret 
the columns of the loading matrix, $W$, as encoding functional connectivity 
networks, the lack of identifiability in PCA and factor analysis models is a 
significant limitation. 
We refer to the model 
presented in \cite{Monti2018} as 
 Modular Hierarchical Analysis (MHA). 
The associated optimization problem therefore becomes:
\begin{equation}
    \label{NNorthoPCA_W}
    \hat W = \argmax_{W: W^TW = I \mbox{ and } W \geq 0 } \left \{  \mathcal{L} \right \}. 
\end{equation}

MHA can therefore been seen to address the two fundamental limitations of 
traditional models such as PCA and factor analysis;
first that the presence of negative values in the loading matrix 
complicates the interpretation of such matrices (addressed via the use of non-negativity
constraints) and second is the fact that 
the latent variables are
rotationally invariant (addressed via the further introduction of orthogonality). Furthermore, from the perspective of fMRI data, 
MHA corresponds to an intuitive generative model whereby latent variables 
capture the activity within each functional network.

Finally, we note that model introduced by \cite{hirayama2016characterizing}, termed 
Modular Connectivity Factorization
(MCF), shares many similarities with MHA. 
In fact, both methods introduce non-negativity and orthonormality over the 
loading matrix, $W$. The fundamental difference, however, is that MCF is not associated with a 
linear latent variable model, and instead parameters are inferred as follows:
\begin{equation}
    \hat W = \argmax_{W: W^TW = I \mbox{ and } W \geq 0 } \left \{  \sum_{i=1}^N \mbox{tr} \left ( {\Sigma^{(i)}} K^{(i)} \right )^2 \right \},
\end{equation}
where $\Sigma^{(i)}$ is defined as in equation (\ref{linLVM_covModel_exapanded}) and $K^{(i)}$ is the empirical 
covariance for the $i$th subject. A related approach was also 
  proposed by \cite{Hyvarinen2016}.

Figure \ref{fig:PCAextensions} provides a 
visualization of the benefits obtained by introducing each of the aforementioned constraints. In particular, we note that it is the combination 
of non-negativity together with orthonormality which yields interpretable, clustered 
networks. 
We  empirically validate such claims by applying all of the aforementioned 
models to synthetic and real fMRI datasets below. 
We discuss the optimization of equations (\ref{PCA_W}), (\ref{NNPCA_W}) and (\ref{NNorthoPCA_W}) in the supplementary material. 


\begin{figure}
    \centering
    \includegraphics[width=\textwidth]{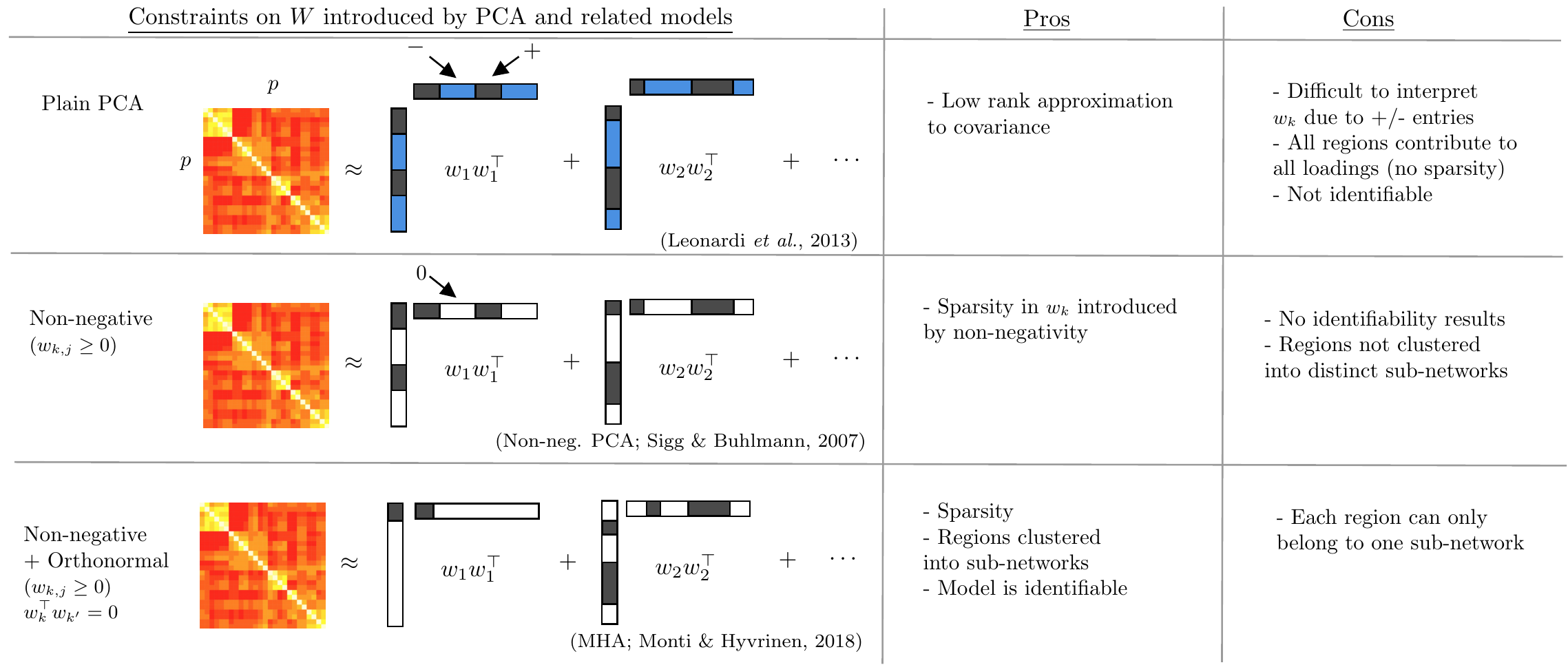}
    \caption{Figure demonstrating the relationship between linear latent variable models, such as PCA and its 
    extensions, to inferred 
    networks. We highlight how introducing various structural constraints on the 
    loading matrix, $W$, improves interpretability of such models. }
    \label{fig:PCAextensions}
\end{figure}

\subsection{Predicting brain age using functional network activity}

The previous section outlined the various flavours of latent variable models which can be employed in order to learn 
functional networks across a cohort of $N$ subjects. 
The aforementioned 
models allow us to decompose 
observed functional connectivity patterns as a linear sum of 
networks encoded by the columns of the loading matrix, $W$.
While the loading matrix is shared across all subjects (indicating the same 
networks are present across all subjects), 
the extent to which they contribute to the observed covariance of  the $i$th subject is denoted by the diagonal entries of 
$G^{(i)}$, as
stated in equation (\ref{linLVM_covModel_exapanded}).

We now consider the task of predicting the biological brain age, $a^{(i)}$, using inferred functional connectivity 
networks as features. 
In the interest of interpretability we limit ourselves to linear regression models of the form:
\begin{equation}
    \label{lin_predict_eq}
    a^{(i)} = \sum_{j=1}^k \beta_j g_j^{(i)} + \epsilon^{(i)}.
\end{equation}
Recall that $g_j^{(i)}$ corresponds to the $j$th diagonal entry of the 
matrix $G^{(i)}$.
As such, the proposed models will essentially seek to predict the biological age of subjects 
by considering activity within each inferred 
functional network. In the case of the $i$th subject, the observed activity in 
network $j$ is quantified by $g_j^{(i)} \in \mathbb{R}_+$. 
In practice, we will seek to quantify the activity of various functional networks on unseen subjects, defined to be 
subjects whose data was not employed to estimate loading matrix, $W$. 
We note that due to the orthonormality of $W$, together with equation (\ref{linLVM_covModel_exapanded}),  
we may estimate $g_j^{(i)}$ for data from unseen subjects, denoted
by $i^*$,  as follows:
\begin{equation}
    \label{estimate_g}
    \hat{g}_j^{(i^*)}  = W_j^T \hat \Sigma^{(i^*)} W_j - v^{(i^*)}. 
\end{equation}
We note that equation (\ref{estimate_g}) requires the observation noise, $v^{(i^*)}$. 
This is not a concern for all subjects whose data is employed during the unsupervised 
learning of the latent variables, as parameters $v^{(i)}$ are inferred alongside loading matrix, $W$. 
However, the primary goal of this work is to build predictive models which can generalize to unseen subjects.
In this context, an estimate of the observation noise, $v^{(i^*)}$, can be obtained as follows:
\begin{equation}
\label{get_noise_unseen}
    \hat v^{(i^*)} = \mbox{tr} ~ \hat \Sigma^{(i^*)} - W^T \hat \Sigma^{(i^*)} W. 
\end{equation}

Although the class of models considered in equation (\ref{lin_predict_eq})  may be considered amongst the simplest supervised regression models,  
they yield 
several important benefits when seeking to understand both the estimated parameters 
as well as the contribution of each of the features.
In particular,  each $\beta_j$ corresponds to the regression coefficient summarizing the (linear) relationship between the activity of 
the $j$th 
network and biological age, conditional on all remaining 
networks. As such, if 
certain regression coefficients are deemed to be insignificant, we may conclude that the 
associated 
network is invariant during healthy ageing.

\subsection{Hyper-parameter selection}
\label{sec::traintestsplit}
The proposed two-stage estimation framework requires the input of only one hyper-parameter: the dimensionality of latent variables $k$. 
In the context of PCA and factor analysis, this hyper-parameter directly corresponds to the number of principal components or 
factors inferred, and a wide literature exists for tuning such a parameter \citep{jolliffe2011principal}. 
One of the advantages of the latent variable models presented in Section \ref{linearLVMreview} is that they each correspond to  
probabilistic models whose likelihood can be directly evaluated. 
As such, a logical choice to tuning hyper-parameter $k$ is to directly 
maximize the log-likelihood over held out data. 

In order to effectively perform hyper-parameter tuning as well as quantify the generalization performance 
of the proposed method, data was split into training, validation and test datasets as follows:
\begin{itemize}
    \item First, a subset of subjects were held out as test data. As such, we obtain two datasets:
    $$ \left \{ X^{(i)}_{1:n}, a^{(i)} \right \}_{i \in S_{train}} ~~\mbox{ and } ~~ \left \{ X^{(i)}_{1:n}, a^{(i)} \right \}_{i \in S_{test}}$$
    where $S_{train}, S_{test} \subset \{1, \ldots, N\} $ denote the non-overlapping sets of
    training and test  subjects respectively. Recall $N$ is the number of 
    subjects present and we write $X^{(i)}_{1:n}$ to denote the $n$ observations available for the $i$th subject. 
    \item Training data is further split into training and validation datasets
    on
    a subject-by-subject basis. 
\end{itemize}
     
Splitting the data in this manner allows for effective hyper-parameter tuning, using training and validation datasets,
as well as for generalization performance to be measured using test dataset which corresponds to unseen
subjects. 

\subsection{Experimental data}


The data employed in this manuscript corresponds to resting-state fMRI data taken from three distinct open-access repositories.
There were small variations in the 
resting state functional MR image acquisition 
for each of the repositories considered: 
CamCAN \citep{taylor2015rn}, Human Connectome Project \citep{van2013wu}, and the ATR Wide Age Range \citep{ogawa2018large}. 
The pre-processing employed on each dataset was as follows:
\begin{itemize}
    \item CamCAN: This dataset was pre-processed by us. Data was motion corrected, spatially smoothed with a 5mm FWHM Gaussian kernel, registered into MNI152 standard space using FLIRT \citep{smith2004advances}
     via a skull-stripped high-resolution T1 image and resampled to 4x4x4mm voxel sizes. Each high resolution T1 image was segmented into grey and white matter and cerebrospinal fluid using SPM Dartel \citep{ashburner2009computational}.
     Mean timecourses for cerebrospinal fluid and white matter as well as 6 motion parameters were linearly filtered from each voxel to reduce non-neural noise.
     
     \item HCP:  We used the pre-processed resting state fMRI data from a random subset of 
     healthy participants\footnote{Full details of the pre-processing pipeline can be found at \url{https://www.humanconnectome.org/study/hcp-young-adult/document/extensively-processed-fmri-data-documentation}}.
     Notably, the pipeline involved FIX ICA-based noise reduction process \citep{salimi2014automatic}, to remove individual sources of physiological, non-physiological and motion related noise.
     
     \item ATR: We used the preprocessed data\footnote{Full details are provided here \url{https://bicr-resource.atr.jp/var/www/webapp/bicrresource/bicrresource/staticfiles/pdf/Methods.pdf}}. 
     The pre-processing pipeline notably included regressing out the global grey matter signal as well as signals from cerebrospinal fluid and white matter, to remove sources of spurious variation.

\end{itemize}

All three pre-processed fMRI datasets were subsequently processed as follows: a cortical parcellation based on resting state functional connectivity analyses \citep{power2011functional} was used to define 264 distinct 10mm diameter regions of interest (ROIs). The fMRI time course averaging across all voxels within each ROI was extracted. These 264 average time courses were then used in subsequent analyses.


\section{Results} 
\label{sec:ExpRes}
In this section we present a range of experimental results involving both synthetic and real resting-state fMRI datasets.
Throughout this section, we contrast the performance 
of the various linear latent variable models presented in Section \ref{linearLVMreview}. 
%
In particular, we study the performance across
the following methods: 
factor analysis (FA), PCA, non-negative PCA \citep{sigg2008expectation},  MCF \citep{hirayama2016characterizing}
and MHA \citep{Monti2018}. 

We first present results using synthetic data in Section \ref{sec::SimStudy}. 
These simulation experiments serve as a numerical validation of the proposed two-stage procedure.
Experiments relating to brain age prediction from resting-state fMRI data are subsequently presented in 
Section \ref{subsec::realdataexp}. 


\subsection{Synthetic data experiments}
\label{sec::SimStudy}
In this section we evaluate the performance of the proposed two-stage estimation framework 
using synthetic data. To this end, 
we generate artificial data whose properties approximately match those which are frequently 
reported in fMRI studies.
The objective is then to quantify which of the linear latent variable models 
presented in Section \ref{linearLVMreview} are able to 
both robustly recover the associated loading matrix, $W$, as well as learn the 
relevant factors which serve as accurate predictors of brain age on unseen subjects. 

Synthetic data was then generated  in order to satisfy equations 
(\ref{pca_gen_1}-\ref{pca_gen_2})  and (\ref{lin_predict_eq}). This is achieved as follows:
\begin{itemize}
    \item 
    First, we randomly generated a factor loading matrix, $W \in \mathbb{R}^{p \times k}$, which satisfied 
    the constraints of both non-negativity and orthonormality. The reason for introducing both  constraints is that we will seek to quantify how reliably each latent variable model can recover $W$, and it is therefore imperative to ensure we generate $W$ from an identifiable model (see discussion in 
    Section \ref{linearLVMreview}). 
    In order to achieve this a dense matrix, $W$, was sampled with each entry 
    following a uniform distribution over the interval $[0,1]$.
    Subsequently, 
    for each row only the entry with the largest value was retained with all other entries set to zero.
    Finally, the norm of each  column was set to one.  
    \item Second, the factor loadings for the $i$th subject, $g^{(i)} \in \mathbb{R}^{k}$,
    were randomly generated as follows:
    $$ g^{(i)}_j \sim \mathcal{N}(2.5, 1.0), ~~ \mbox{for $j=1, \ldots, k$} $$
    with all negative samples being discarded. 

    \item 
    The regression coefficients, $\beta \in \mathbb{R}^k$, were  drawn uniformly at random 
    from the interval [0,10].
    
    \item Finally, we are able to randomly generate 
    observations and ages for each subject as follows:
    \begin{align}
        X^{(i)} &\sim \mathcal{N}(0, W G^{(i)} W^T + v^{(i)}), \\
        a^{(i)} &\sim \mathcal{N}( \beta^T g^{(i)} , \epsilon  ).
    \end{align}
    Recall that $G^{(i)} \in \mathbb{R}^{k \times k}$ is a diagonal matrix 
    consisting of entries $g^{(i)}_j$. 

\end{itemize}

We note that the choices for sampling distributions of both the 
factor loadings, $g^{(i)}$, as well as the regression coefficients, $\beta$, are
necessarily somewhat heuristic. However, care was taken to ensure the 
implied distributions over subject ages approximately matched the empirical distributions observed within the
camCAN repository. 

We note that throughout experiments we consider the performance of each method 
whilst varying two distinct factors:
the number of observations per subject, $n$, and the number of 
training subjects, $N$. Furthermore, throughout simulations we fix the dimensionality of observations to 
be $p=50$ and the number latent factors to be $k=5$. 



Given artificial data generated as described above, we look to quantify 
the performance of each of the  linear latent variable models using the following two metrics:
\begin{enumerate}
    \item 
    Accurate recovery of the loading matrix, $W$. This is quantified in terms of the
    squared error between the true loading matrix and the estimated loading matrix.
    \item Accurate brain age prediction over unseen subjects. In line with other literature, this 
    is quantified in terms of the mean absolute error between true and predicted brain ages \cite{franke2010estimating, lancaster2018bayesian}. 
\end{enumerate}

\subsubsection{Synthetic data results}

\begin{figure}[h!]
\centering
\includegraphics[width=\columnwidth]{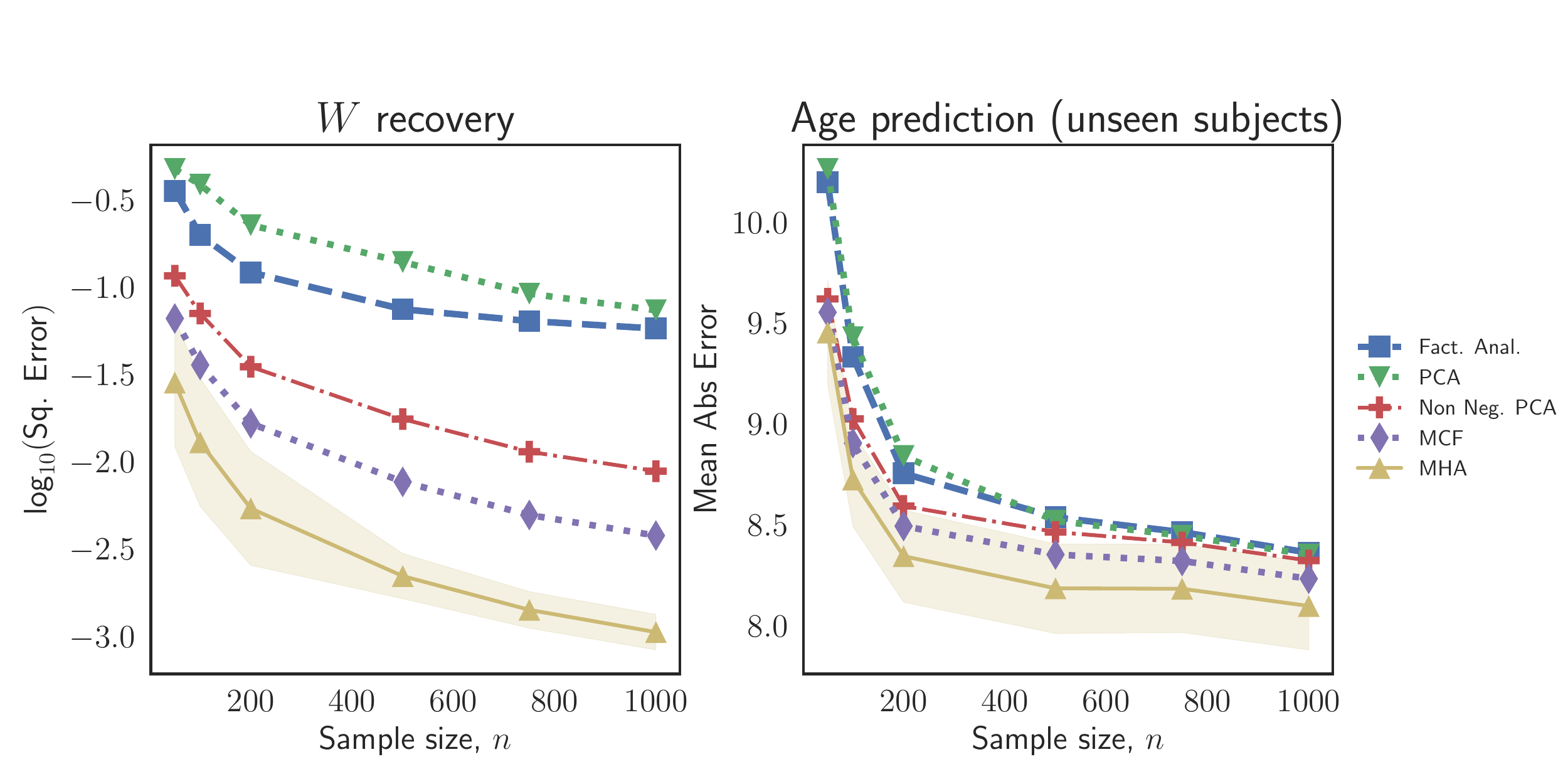}
\caption{Simulation results for recovery of the true loading matrix (left panel) and 
prediction of brain age for unseen subjects (right panel) as the number of 
observations per subject, $n$, increases. We note that the introduction of 
regularity constraints (e.g., non-negativity or orthonormality) on the loading matrix 
leads to improvement in performance. }
\label{SimRes_incObs}
\end{figure}

We begin by considering the performance of each linear latent variable model as the number of 
observations per subject, $n$,  increases for a fixed number of training subjects, $N=25$. The results are presented in 
Figure \ref{SimRes_incObs}. We note that both in terms of recovery of the loading matrix, $W$, as well
as in terms predicting the ages over unseen subjects, the introduction of regularity constraints (be they in the 
form of non-negativity or orthonormality) leads to improvements. 
This is not entirely surprisingly as the true loading matrix in these experiments satisfies these 
conditions. However, it is important to note that even as the 
number of observations, $n$, increases significantly methods such as PCA and factor analysis continue to 
perform less competitively than methods which introduce regularity constraints. This is a phenomenon is also 
observed in the real data analysis. 

\begin{figure}[t!]
\centering
\includegraphics[width=\columnwidth]{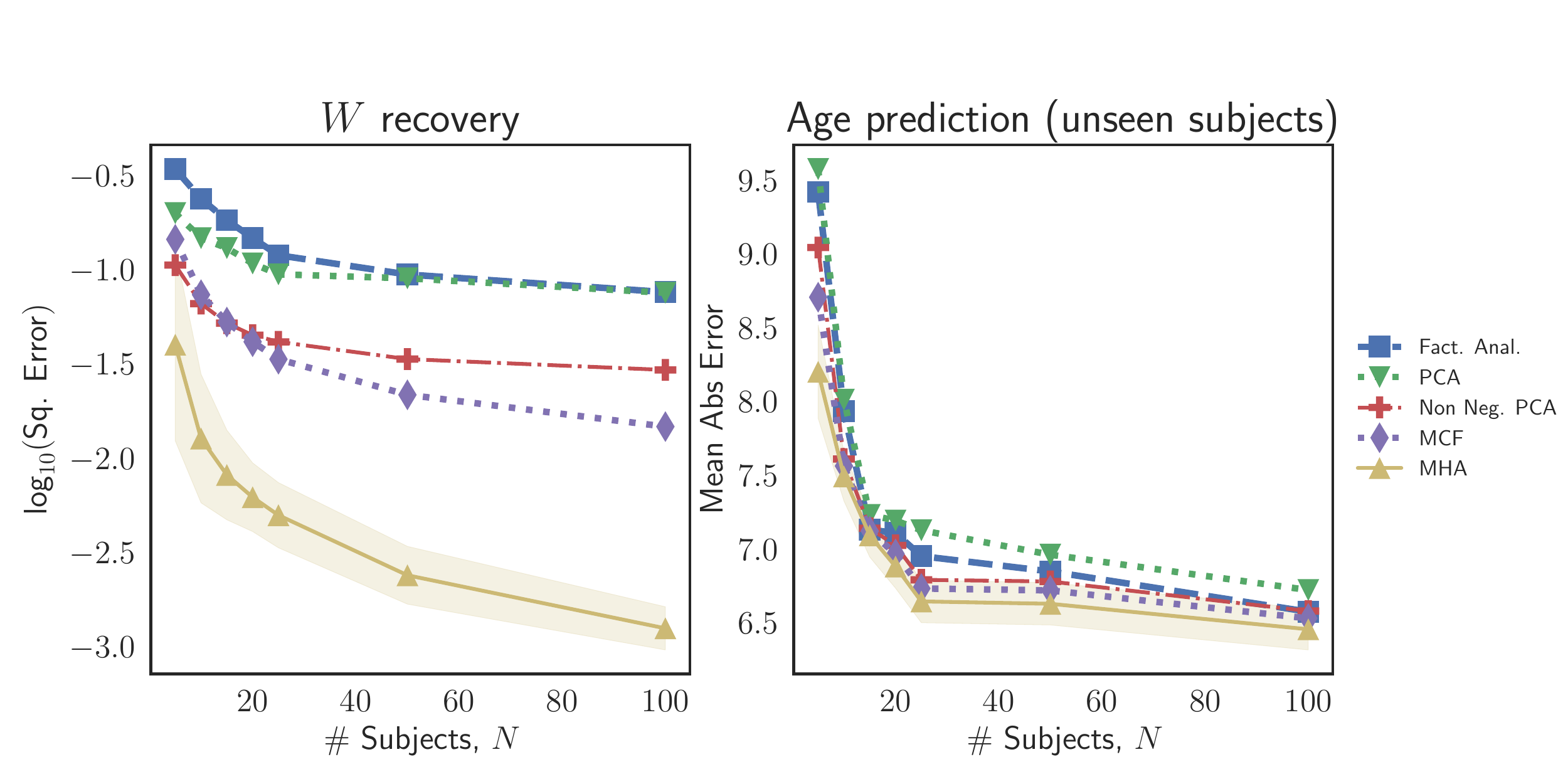}
\caption{Simulation results for recovery of the true loading matrix (left panel) and 
prediction of brain age for unseen subjects (right panel) as the number of 
training subjects, $N$, increases. We note that the introduction of 
regularity constraints (e.g., non-negativity or orthonormality) on the loading matrix 
leads to improvement in performance.}
\label{SimRes_incSubs}
\end{figure}

We also study the performance of the various latent variable models
when the number of training subjects, $N$, increases and 
the number of observations  is fixed at $n=100$ per subject. 
These results are presented in Figure \ref{SimRes_incSubs}. 
In terms of recovery of the loading matrix, $W$, we again observe that 
introducing regularity constraints leads to significant improvements. 
In terms of predictions over unseen subjects (as shown in the right panel of Figure \ref{SimRes_incSubs}),
the improvements due to the introduction of regularity conditions begin to fade as the number of 
training subjects increases. In particular, beyond a certain number of training subjects (approximately 25 in the case of these experiments), the improvement in out-of-sample predictions begins to plateau. 

\subsection{Resting-state fMRI data experiments}
\label{subsec::realdataexp}

While the previous section 
presented results relating to synthetic data, here 
we present experimental results where the 
proposed two-step procedure is applied to three open-source resting-state fMRI datasets. 
The datasets considered correspond to the Cambridge Center for Ageing and Neuroscience (CamCAN) repository, 
the Human Connectome Project (HCP) repository, 
and the ATR Wide-Age-Range  repository. 
The purpose of employing three distinct datasets is to effectively measure 
the generalization performance of the proposed approach on unseen data. 
As such, data from the HCP and Wide-Age-Range repositories was not employed during 
any of the model training and instead used exclusively as unseen test data. It is important to note  
that in addition to significant inter-subject variability \citep{kelly2012characterizing}, 
fMRI data also suffers from the presence of several other 
well-documented issues such as variable scanner performance or noise 
\citep{friedman2006reducing,bennett2010reliable, poldrack2011handbook}.
As such, validating the performance of  the proposed 
brain age prediction models in this way 
will provide a more realistic measure of their generalization performance.

\subsubsection{CamCAN repository results}
\label{subsec:camcanres}

Resting-state fMRI data was collected from a total of 647 subjects from the 
CamCAN repository. 
Subject ages ranged from 18 to 88 years of age (average age of 54.31$\pm$18.56, 318 males and 329 females).
%
The CamCAN dataset was employed as the principal dataset in the proposed two-step procedure, implying that it was employed to
learn both the 
functional network structure in the unsupervised learning stage and 
the linear regression models in the supervised learning stage.
As such, the data was split into training, validation and test subsets as described in Section \ref{sec::traintestsplit}.

\subsubsection*{Step 1: unsupervised functional network inference}

The first stage of the proposed framework involves the estimation of 
reproducible functional connectivity 
networks via the use of the various linear latent variable models 
discussed in Section \ref{linearLVMreview}. 
%
The number of 
functional networks inferred corresponds directly to the dimensionality of latent variables,
which is determined by hyper-parameter $k$. As each linear latent variable model can be interpreted as a 
probabilistic model, we select hyper-parameter $k$ by maximizing the log-likelihood over the validation dataset.
This resulted in the choice of $k=5$ when the loading matrix was restricted to be 
both non-negative and orthonormal, as proposed by \cite{hirayama2016characterizing} and \cite{Monti2018}.  
While it is possible
that the choice of hyper-parameter may vary across distinct latent variable models (e.g., for PCA or factor analysis), 
we choose to keep the choice of $k$ fixed across all models as this facilitates model comparison and interpretation of results.

\begin{figure*}[t!]
	\centering
	\includegraphics[width=.8\textwidth]{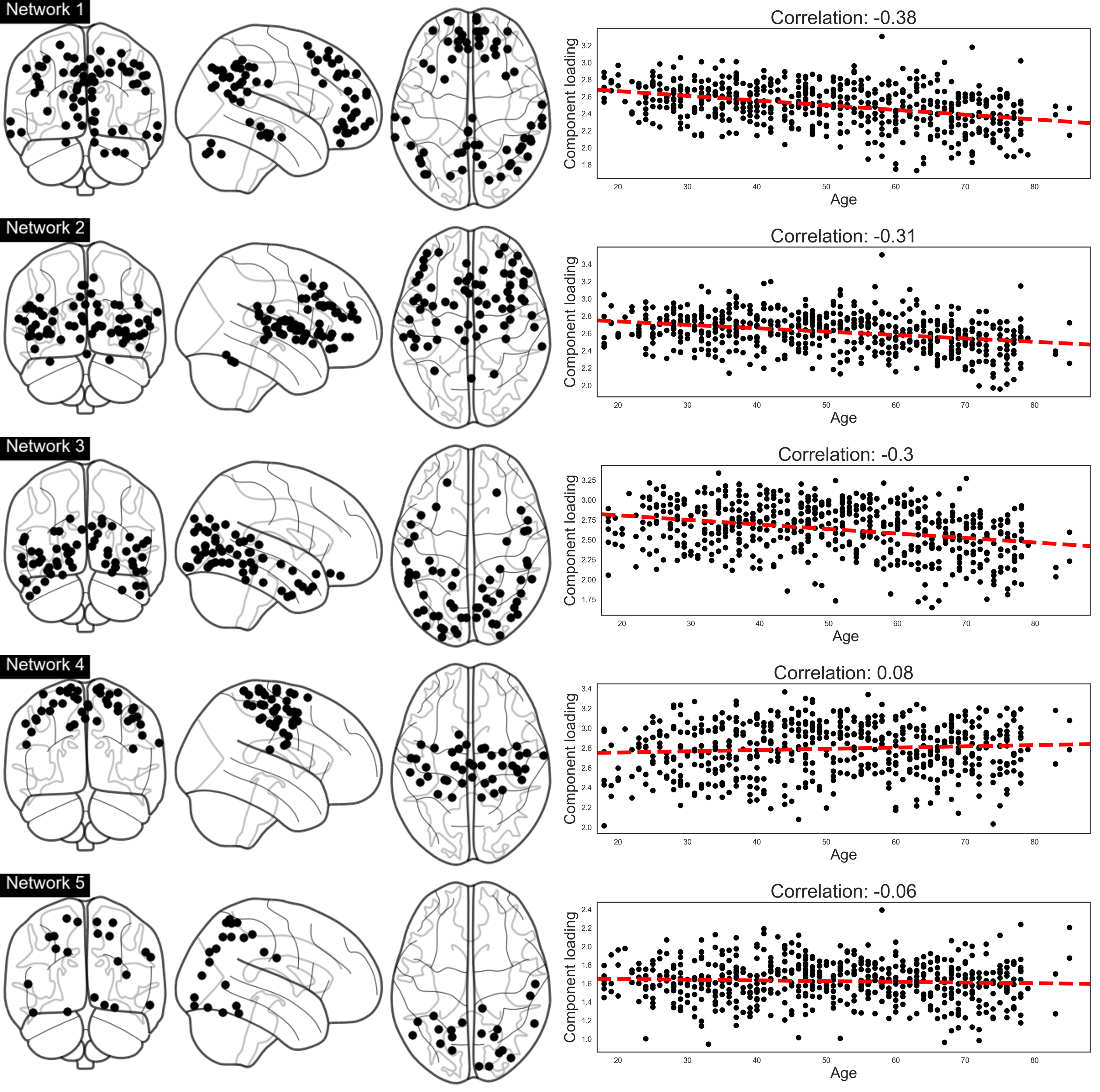}
	\caption{Left panel: inferred networks as recovered when 
		non-negativity and orthonormality constraints are introduced over the loading matrix, $W$. Networks are 
		spatially consistent and symmetric. Right panel: visualization of network activities against 
		subject age demonstrating (mostly negative) linear trends with healthy ageing. }
	\label{camCAN_networks}
\end{figure*}

The left panel of Figure \ref{camCAN_networks} visualises the results when the MHA linear latent variable model was employed. The results demonstrate that the inferred networks 
are spatially homogeneous and symmetric across both hemispheres. Furthermore,
many of the inferred networks  correspond to 
widely reported networks and regions:
network 1 captures the default model network (DMN) and network 2 overlaps with the salience
network, while networks 3 and 4 correspond to a higher-level visual network and the somatomotor
network respectively.
For comparison, we include equivalent plots for all other 
latent variable models considered in  \ref{networks_altMethods}. 
We note that alternative methods, such as PCA, which did not enforce the combination of both
non-negativity and orthonormality, yielded results which were visibly less clustered and 
more difficult to interpret.

The right panel of Figure \ref{camCAN_networks} visualizes the  correlation between the activity of 
each network 
(as defined in equation (\ref{estimate_g})) with the age of each subject.
For  networks 1-3 we observe a significant negative correlation between the activity and age, suggesting that 
ageing induces a drop in activity of such networks. 
These results are in line with related research on ageing induced differences in 
functional connectivity. In particular, the decrease in activity of the DMN (network 1), 
has been widely reported \citep{geerligs2015state, grady2016age, liem_geerligs_damoiseaux_margulies_2019}. 


\subsubsection*{Step 2: supervised training of brain age prediction models }

Recall that the overall objective of the proposed 
framework was build interpretable 
models of biological brain age. To this end, the features recovered from linear latent variable models 
where employed as features in a linear regression framework to predict the brain age of each subject.
In particular, the five distinct  the linear latent variable models detailed in Section \ref{linearLVMreview}
where employed to learn 
reproducible sub-networks parameterized by a loading matrix, $W \in \mathbb{R}^{p \times k}$. 
The activity within each 
functional network, defined as in equation (\ref{estimate_g}), was subsequently employed as features
to predict biological age using linear regression. 


\begin{figure*}[t!]
\centering
\includegraphics[width=.65\textwidth]{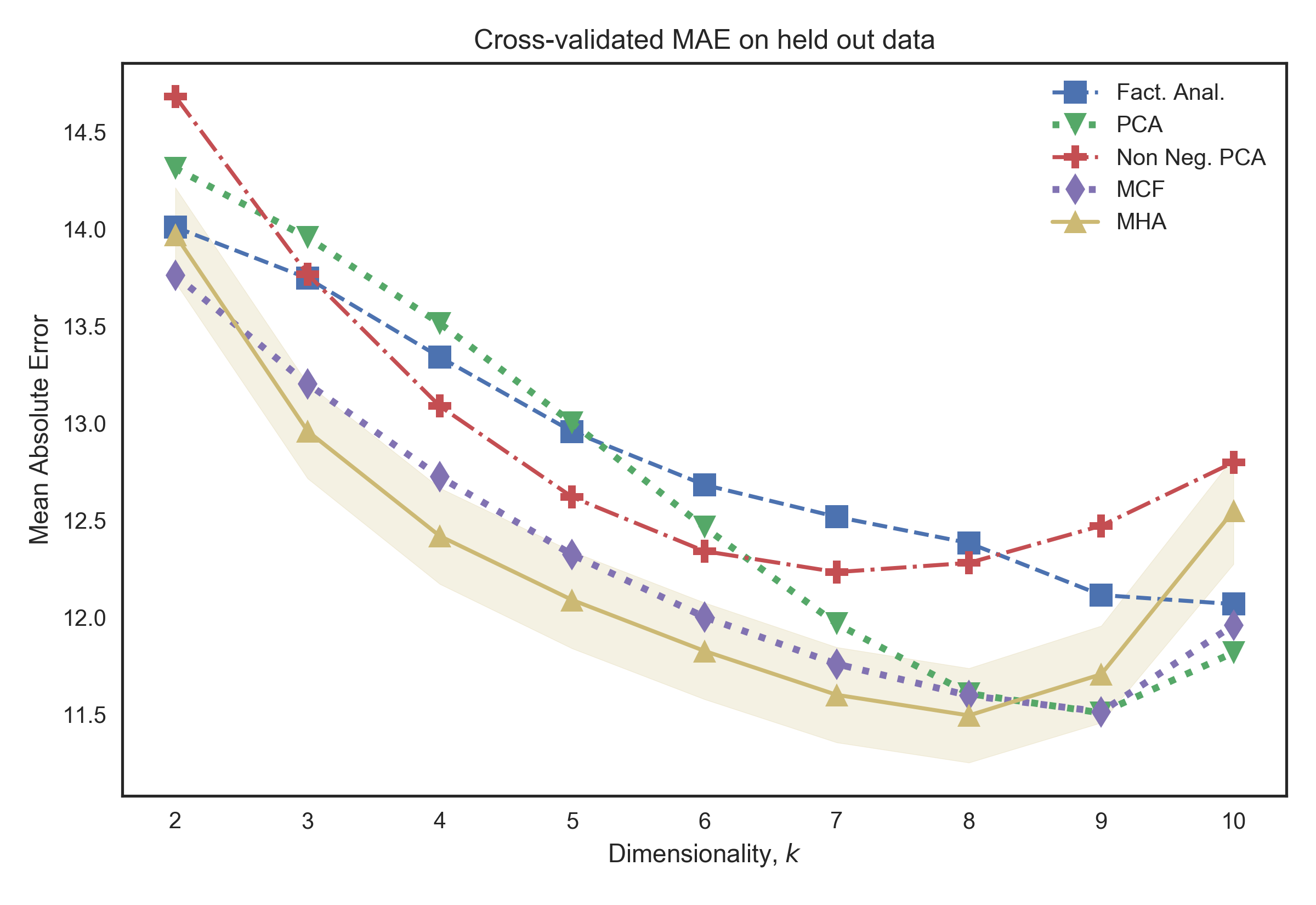}
\caption{Mean absolute error (MAE) performance 
for a varying number of 
networks, as determined by $k$ (x-axis), on unseen test data from camCAN. 
We note that the combination of non-negativity and othonormality (MHA and MCF) yields competitive results across a wide range of $k$.
}
\label{CV_MAE_camcan}
\end{figure*}

We note that the CamCAN repository, as well as HCP and ATR repositories, each contained over 
a hundred subjects each. This is in contrast to typical fMRI studies, where the sample size is often in the 
range of 20 to 30 subjects \citep{poldrack2011handbook, cremers2017relation}. 
Furthermore, recall that the
goal of experiments presented are to quantify performance on unseen resting-state 
fMRI data with a view to providing an 
indication of how each of the linear latent variable models employed would perform in a typical fMRI study.
As such, throughout the remainder of this section we report the performance, in terms of mean absolute error, over random subsets of 30 subjects from each repository. This corresponds to a form of bootstrapping, where we average results over a random sample of possible \textit{cohorts}. In practice, we report results over 1000 random subsets of 30 subjects for each of the 
three repositories considered. 


Figure \ref{CV_MAE_camcan} visualizes the mean absolute error on unseen test data for various choices of 
$k \in \{2, \ldots, 10\}$ . We note that the combination of linear regression with the use of 
non-negativity and orthonormality constraints, as advocated by both 
the MCF and MHA  models 
leads to competitive performance over 
a range of choices of $k$. In particular, such algorithms out-perform both non-negative PCA and PCA, suggesting 
that the introduction of such constraints serves to improve the predictive properties of the model. 
Furthermore,  Figure \ref{CV_MAE_camcan} 
indicates the presence of a bias-variance 
trade-off that is often encountered in supervised learning whereby performance on unseen test data
begins to deteriorate as the number of parameters (in our case $k$) increases beyond a certain value.

\begin{figure*}[b!]
\centering
\includegraphics[width=\textwidth]{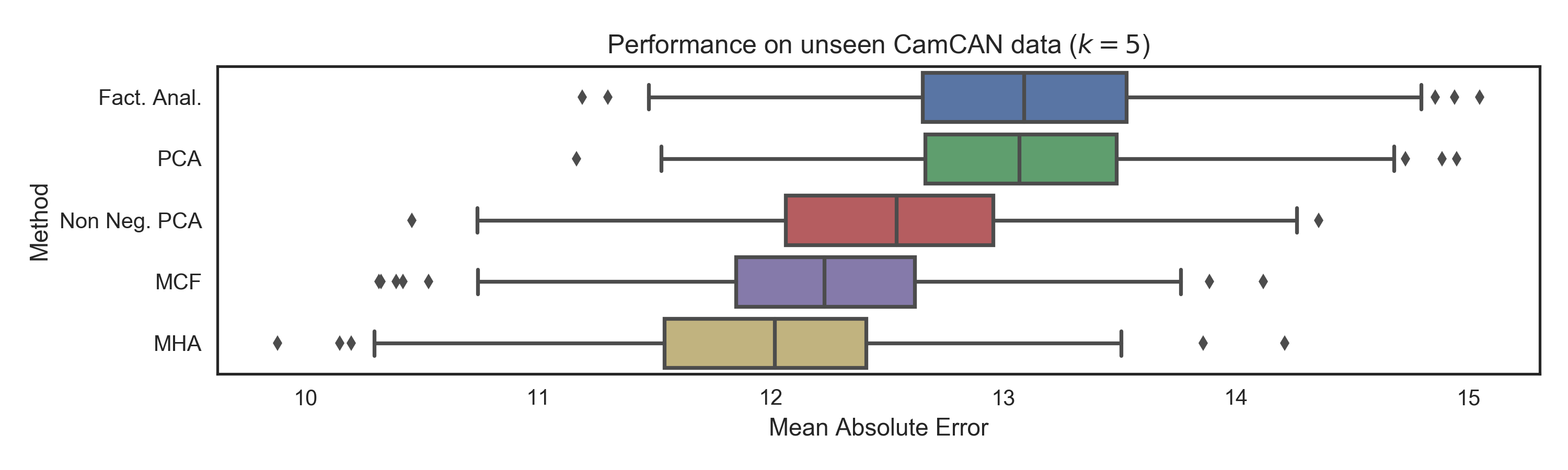}
\caption{
Mean absolute error (MAE) performance on unseen testing data from CamCAN repository when the 
dimensionality of latent variables is fixed to 
$k=5$ (implying we infer 5 networks). 
We note that as regularity constraints are introduced, in particular non-negativity and orthonormality,
predictive performance improves. 
}
\label{CV_camcan_k5}
\end{figure*}

As mentioned previously, the choice of $k=5$ was selected in by maximizing log-likelihood over a validation dataset (i.e., in an 
entirely unsupervised manner - data regarding subject ages was not considered). 
Figure \ref{CV_camcan_k5} visualizes the performance on the unseen test dataset for the specific choice of $k=5$, for all
possible choices of linear latent variable models. The results indicate that as additional constraints are 
introduced to the loading matrix, the generalization capabilities of the models also improve. As such, MCF and  MHA, 
which introduce the most stringent constraints corresponding to \textit{both} non-negativity and 
orthonormality, obtain the best generalization performance. 
Thereafter non-negative PCA, which relaxes the requirement for 
orthonormality, is the next most competitive latent variable model. Finally, PCA and factor analysis, which relax all the 
aforementioned constraints, obtain the worst generalization performance. 
Figure \ref{CV_MAE_camcan} further suggests that this pattern approximately holds for a wide range of $k$.  


\subsubsection{Transfer onto HCP and ATR Wide-Age-Range repositories}
The results of Section \ref{subsec:camcanres} provide a measure of 
performance, in terms mean absolute error in predicted brain age, 
within a large-scale 
resting-state fMRI dataset. 
However, it is widely accepted that in addition subject-specific noise, 
there are several other significant contributors to noise in fMRI data: these include issues related to 
scanner noise and frequency of acquisition of images \citep{friedman2006reducing, bennett2010reliable, poldrack2011handbook}.
As a result, in order to thoroughly verify the generalization performance of the 
proposed methods, we employ resting-state fMRI data from the HCP and ATR Wide-Age-Range repositories. 
We note that data from the aforementioned repositories was employed 
only for testing purposes, as such it was not employed to learn the 
network structure across subjects, nor to tune the parameters of the linear regression models.

\begin{figure}[t!]
\centering
\includegraphics[width=\columnwidth]{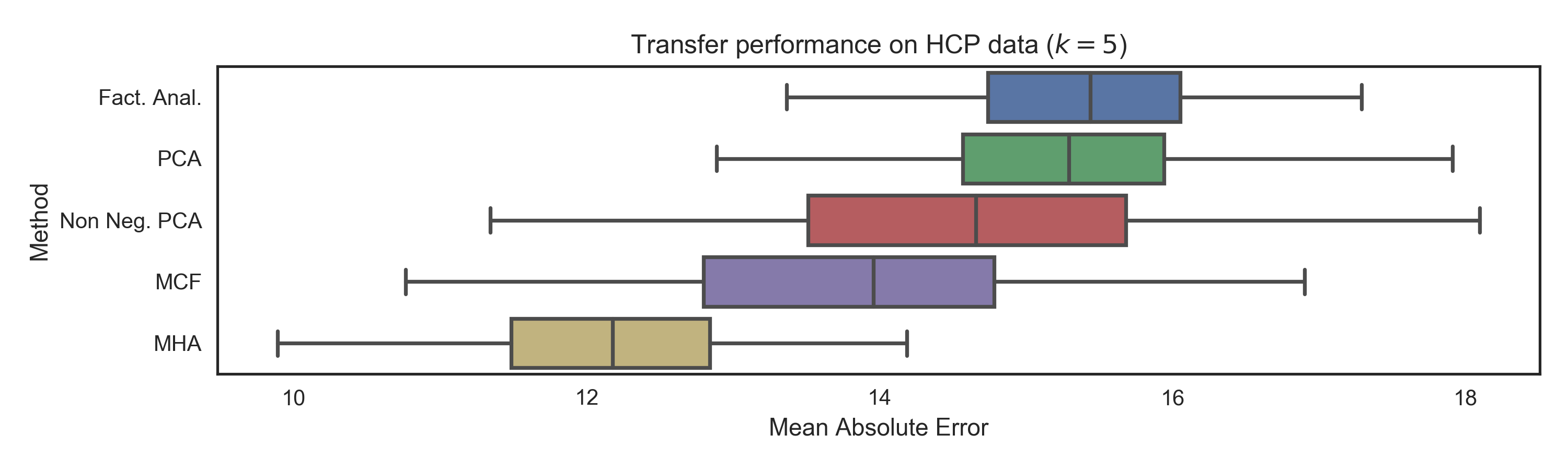}
\caption{
Mean absolute error (MAE) performance on unseen  data from HCP repository. 
Results are broadly consistent with performance on the camCAN data, indicating good generalization. We note that 
the introduction of non-negativity or orthogonality constraints leads to improved generalization (see Table \ref{tab:my_label}).
The number of functional 
networks was $k=5$.
}
\label{MAE_HCP}
\end{figure}   

To summarize, prediction of biological age on both the HCP and ATR Wide-Age-Range repositories  was performed as follows:
First, the loading matrix, $\hat W$ was employed to obtain estimated activity within each  
network, as 
detailed in equations (\ref{estimate_g}) and (\ref{get_noise_unseen}). Subsequently, 
predictions of biological age were obtained using equation (\ref{lin_predict_eq}). 
At each stage both $\hat W$ and $\hat \beta$ are the parameters inferred using the CamCAN dataset (i.e., there 
was no fine-tuning of parameters). As a result, performance on both 
HCP and ATR Wide-Age-Range datasets provide a robust measure of generalization performance to entirely unseen data.

Results on the HCP data are provided in Figure \ref{MAE_HCP}. As expected, the mean absolute errors are larger for each of the 
distinct latent variable models when compared to the results of on the camCAN dataset (Figure \ref{CV_camcan_k5}),
which will be partially the result of varying scanner noise and image acquisition properties. Importantly we note that, as with the 
camCAN dataset, there once again a relationship between the introduction of additional constaints (in the form 
of non-negativity and orthonormality) and generalization performance. As before, methods such as PCA and factor analysis which do not 
introduce any constraints had the weakest performance as well as the largest drop in performance, see Table \ref{tab:my_label}. 
Whereas the methods introducing both non-negativity and orthonormality yielded the best generalization performance as well as the 
smallest drop. These results thereby serve as additional evidence that the introduction of robust constraints can 
serve to improve both the interpretability of results (as discussed in Section \ref{subsec:camcanres}) as well as the predictive power of
associated models.

\begin{figure}[t!]
\centering
\includegraphics[width=\columnwidth]{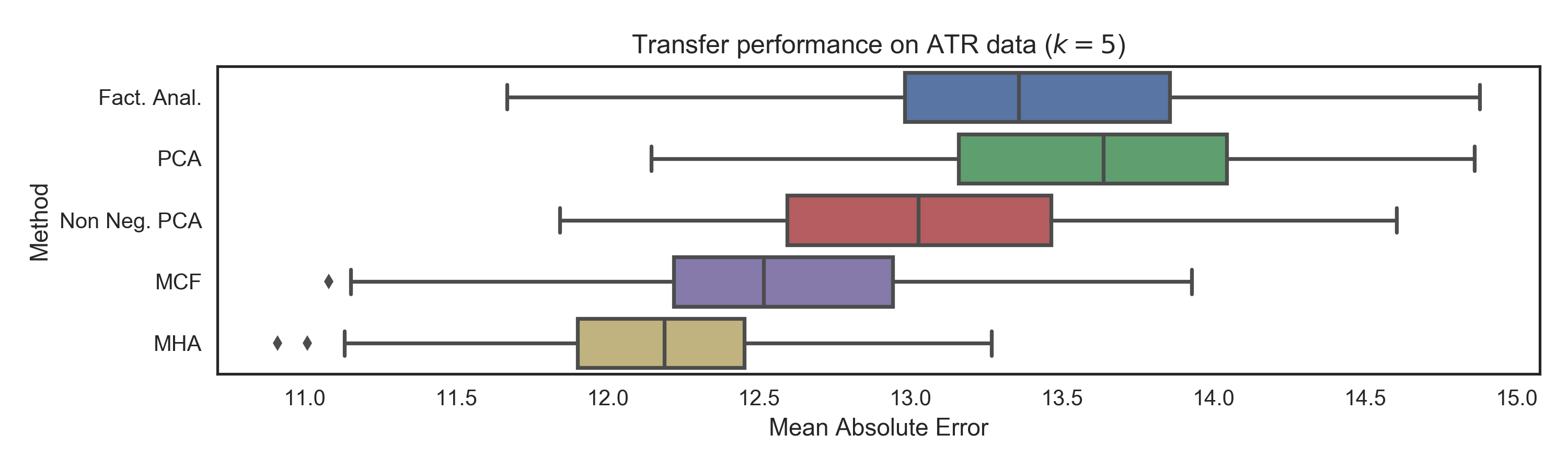}
\caption{Mean absolute error (MAE) performance on unseen  data from ATR Wide-Age-Range repository. 
Results are broadly consistent with performance on the camCAN  data, indicating good generalization. Further, as with the 
HCP data, we note that 
the introduction of non-negativity or orthogonality constraints leads to improved generalization (see Table \ref{tab:my_label}).
The number of functional  
networks was $k=5$.
 }
\label{MAE_ATR}
\end{figure}

The HCP results presented above
serve to partially validate the 
predictive models trained using the camCAN dataset. However, one significant limitation of the HCP dataset is that subject 
ages only range from 22 to 37 years of age. This is particularly relevant in the context of brain age biomarkers, as many neurodegenerative diseases of interest will be associated with advanced ages. As a result, we further validated the 
generalization capabilities of the proposed brain age prediction models on the 
ATR Wide-Age-Range dataset, which had subjects ranging from 20 to 70 years of age. 
Results, presented in Figure \ref{MAE_ATR} and Table \ref{tab:my_label}, are consistent with results on the camCAN and HCP datasets, 
again indicating that the introduction of 
constraints non-negativity and orthonormality constraints improves generalization performance.

\begin{table}[hb!]
\begin{center}
 \begin{tabular}{||l| l l l||} 
 \hline
 Latent var. model &  CamCAN  &  HCP & ATR  \\ [0.5ex] 
 \hline\hline
  Factor analysis & 13.09 (0.62)  & 15.38 (0.92) & 13.40 (0.89)\\
 \hline
   PCA & 13.08 (0.60) &  15.30 (1.09) &  13.65 (0.89)\\
 \hline
 Non-neg. PCA &  12.51 (0.64) &14.61 (1.47) & 13.05 (0.92)\\
 \hline 
 MCF  & 12.23 (0.58) &   13.79 (1.42) &12.59 (0.83)\\ 
 \hline
 MHA &  11.98 (0.62)  & 12.71 (1.01) & 12.18 (0.78)\\ 
 \hline
\end{tabular}
\end{center}
    \caption{Mean absolute error (MAE) performance 
    for various choices of linear latent variable models on 
    each of the three repositories considered (standard deviations in brackets). 
    Latent variable models are ordered according to constraints enforced on the loading matrix: Factor analysis and PCA 
    introduce no constraints, non-negative PCA enforces non-negativity while the remaining models 
    also require orthonormality. 
    %
    }
    \label{tab:my_label}
\end{table}



    


\section{Conclusion} 

It is widely accepted that ageing has pronounced effects on 
the functional architecture of the human brain \citep{geerligs2014brain, smith2019estimation}. 
In the current study we have presented and validated a two-stage framework through which to 
train interpretable and robust models of biological brain age based on
functional connectivity. In particular, the proposed framework first employs 
linear latent variable models to uncover reproducible 
networks which are present throughout a cohort of subjects. 
A variety of such latent variable models are considered many of which
extend PCA by introducing 
constraints such as non-negativity over the loading matrix.
Our experiments suggest that whilst PCA is a natural candidate for dimensionality reduction, and can be interpreted as
recovering latent \textit{eigenconnectivities}, 
the introduction of constraints such as non-negativity can serve to greatly improve both interpretability and 
predictive performance. 

Given inferred functional networks and their activations we train linear predictive models
of biological brain age where in the interest of interpretability we deliberately restrict ourselves to 
linear models. This allows us to directly interrogate the effects of each functional network on the predicted brain age (as shown in  
Figure \ref{camCAN_networks}). In line with other results in the literature, we find a decrease in activation in the 
default mode network, salience network and higher-level visual network as biological age increases. 

The proposed two-stage framework is first validated on the data from the CamCAN repository and subsequently 
further applied to two further open-access repositories: the HCP and ATR Wide-Age-Range 
repositories. 
The use of data from two additional repositories serves to provide a clear empirical indication of the 
generalization capabilities of the proposed approach. This is especially relevant in the context of fMRI data, where
artefacts such as scanner noise can often cause significant challenges \citep{poldrack2011handbook}. 

 
We note that the brain age prediction errors presented in this 
work are not competitive with alternative methods which are based on alternative imaging modalities, such as 
structural imaging data \citep{cole2017predicting_CNN, cole2017predicting}. This is to be expected for two reasons.
First, the imaging modality employed in this work, resting state fMRI data,  
is both noiser and likely to be less age-indicative than 
structural measures. Second, in this work we deliberately restrict ourselves to 
building simple yet interpretable models of brain age. As such, we restrict ourselves to consider 
only linear classifiers as these allow for clear model interpretation and interrogation, while noting 
that the use of more expressive models (e.g., nonlinear models) in the second stage should naturally lead to improved performance. 

Furthermore, it is important to note that whilst this work demonstrates the feasibility of functional connectivity driven models 
of biological brain age, all subjects included in these studies 
were healthy. As such, whilst such models could eventually be employed to develop biomarkers, further 
experimentation and validation will be required in future. 
Moreover, an avenue for further research would be to consider 
performing classification instead of regression in the second stage of the proposed method. Whilst a natural task would be to 
discriminate between healthy controls and subjects with some neuropathology, such an approach could also be employed in 
the context of task-based fMRI. In particular, task-based fMRI has been widely reported as displaying non-stationary 
functional connectivity structure
\citep{monti2014estimating, monti2017real, calhoun2014chronnectome}. As such, seeking to discriminate between various cognitive tasks, 
for example as considered by \cite{monti2017decoding}, could be 
an exciting future application. 
Finally, while in this work we have considered linear latent variable models such as PCA, future
work could consider alternative 
latent variable modes such as latent position graphs 
 \citep{athreya2017statistical}. 


\section*{Acknowledgments} 
\noindent{The authors with to thank Steve Smith for valuable feedback and discussions. }

\newpage 
\bibliographystyle{plainnat}
\bibliography{clean.bib}

\newpage 
\setcounter{section}{0}
\setcounter{figure}{0}
\appendix

\section{Further analysis and validation}

\subsection{Age distributions of subjects across repositories} 
\begin{figure}[h!]
\centering
\includegraphics[width=.7\textwidth]{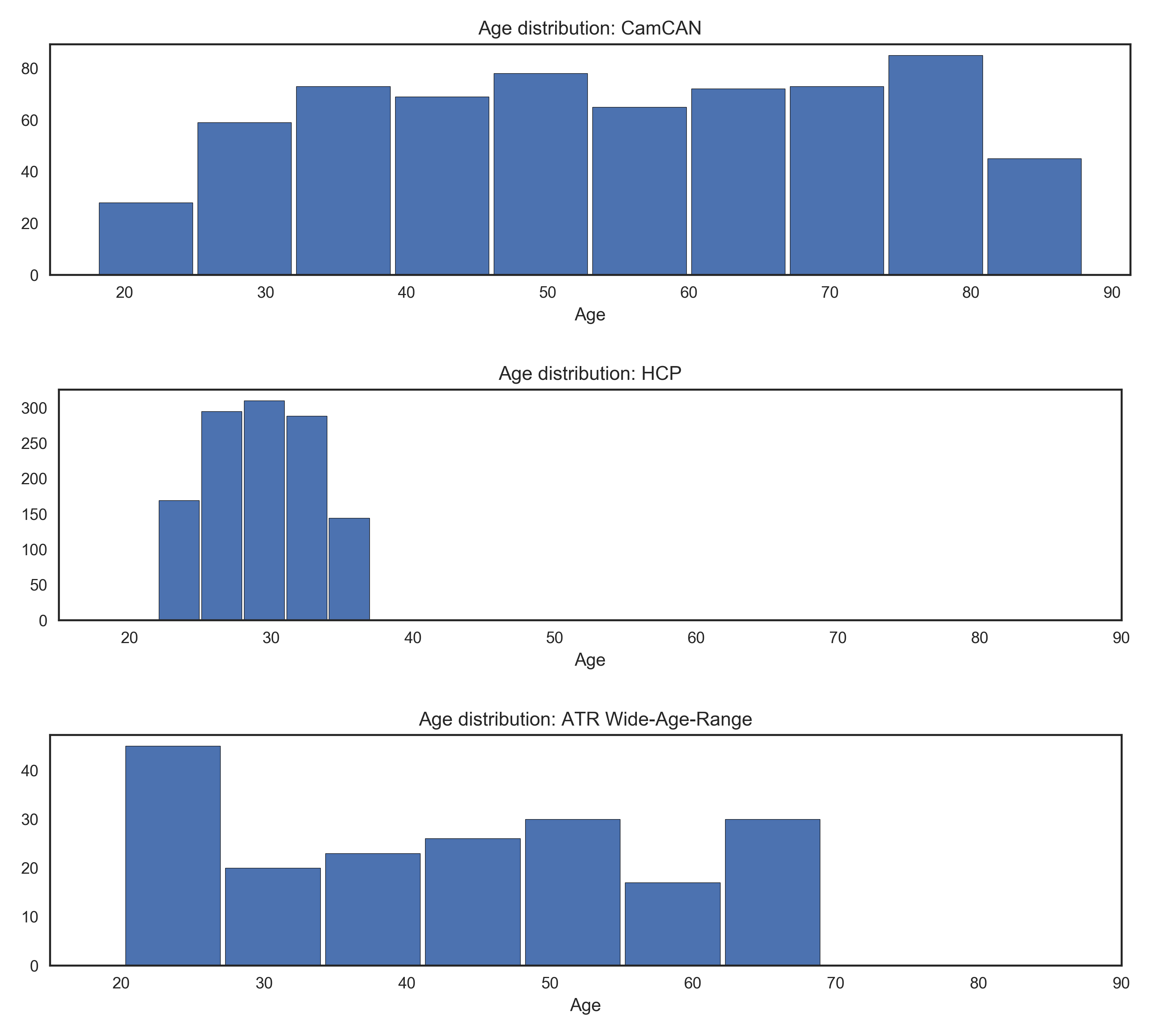}
\caption{Histogram visualizing age distribution for each of the repositories employed. We note that the CamCAN dataset has the widest range of all 
repositories considered, validating its use as a the primary dataset in our study. }
\label{ageHistCamCAN}
\end{figure}


\newpage
\section{Functional connectivity networks inferred by PCA and alternative models}
\label{networks_altMethods}

\begin{figure*}[!ht]
\centering
\includegraphics[width=.6\textwidth]{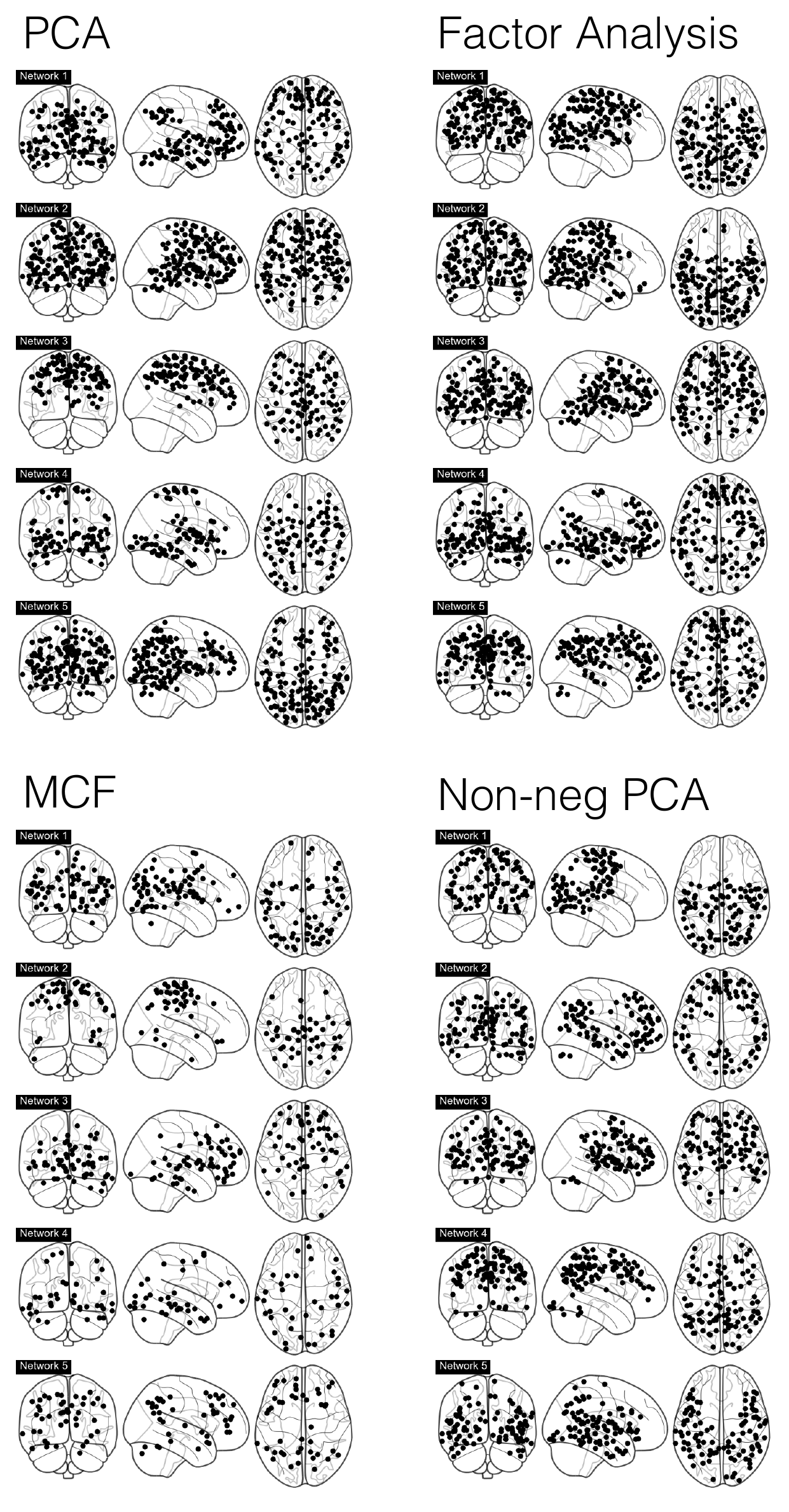}
\caption{Inferred networks using alternative linear latent variable models. In the case of models such as PCA and factor analysis, networks  were obtained by 
thresholding entries of $W$ so only non-negative entries considered.}
\label{lab:PCAnet}
\end{figure*}

\clearpage
\newpage
\section{Technical details of our methodology}\label{appA}





In this appendix, we give further details of the block-coordinate descent algorithm which we implement to update the model parameters.
In practice, we solve the constrained optimisation \eqref{NNorthoPCA_W} via the use of projections onto the non-negative quadrant (non-negativity) and Lagrange multipliers.  More specifically, we use the objective function:
\begin{equation}
    \tilde{\mathcal{L}} = \mathcal{L} + \frac{\delta}{2} || W^T W - I_k ||_2^2 + \mbox{tr} (\Gamma^T (W^T W - I_k)) ,
\end{equation}
where $\Gamma \in \mathbb{R}^{k \times k}$ and $\delta$ are Lagrange multipliers enforcing the orthonormality constraints.


We employ gradient descent approach to update the estimate of $W$.  To this end, we follow \cite{Monti2018} and introduce a gradient step size $\eta$ and project onto the non-negative orthant at each iteration (this ensures that the positivity constraint is maintained).  The update takes the form
\begin{equation}\label{Wupdate}
W \leftarrow \mathcal{P}^{+} \left(W - \eta \left(\frac{\partial \mathcal{L}}{\partial W } +\delta (W W^{\top} W - W) +W \Gamma) \right)\right), 
\end{equation}
where $\mathcal{P}^{+}=\operatorname{max}(0,x)$ denotes the projection onto the non-negative orthant and $\eta$ is a stepsize parameter. 
The update for the Lagrange multipliers $\Gamma$ is given by \citep{Bertsekas2014}:
$$
\Gamma \leftarrow \Gamma + \delta(W^{\top} W - I).
$$




In the case of the loading matrix, the gradient update is defined as:
\begin{align}
\label{ML_wupdate}
\frac{\partial \mathcal{L}}{\partial W } &= \sum_{i=1}^N \frac{\partial \mathcal{L}}{\partial \Sigma^{(i)}}   \frac{\partial \Sigma^{(i)} }{\partial W } \\
&= \sum_{i=1}^N \left (  - {\Sigma^{(i)}}^{-1} + {\Sigma^{(i)}}^{-1} S^{(i)} {\Sigma^{(i)}}^{-1} \right ) W G^{(i)}\;,\nonumber
\end{align}
where we note that via the Sherman-Woodbury identity and using the form of the covariance \eqref{linLVM_covModel}, we can write ${\Sigma^{(i)}}^{-1}$ as
follows:
\begin{equation}\label{ShermanWoodbury}
    {\Sigma^{(i)}}^{-1} = {(v^{(i)}I)}^{-1} - {(v^{(i)}I)}^{-1}  W ( {G^{(i)}}^{-1} + W^{\top} {v^{(i)}I} W)^{-1}W^{\top} {(v^{(i)}I)}^{-1}.
\end{equation}
For the diagonal matrix of eigenvalues, $G^{(i)}$, we can update each matrix independently as follows:
\begin{align}
\frac{\partial \mathcal{L}}{\partial G^{(i)} } &=  \frac{\partial \mathcal{L}}{\partial \Sigma^{(i)}}   \frac{\partial \Sigma^{(i)} }{\partial G^{(i)} }
\\
&=  \sum_{i=1}^N \left ( - {\Sigma^{(i)}}^{-1} + {\Sigma^{(i)}}^{-1} K^{(i)} {\Sigma^{(i)}}^{-1} \right ) \;. \nonumber 
\label{Lambda_mlupdate}
\end{align}

\end{document}